# On the Black Hole's Thermodynamics

# and the Entropic Origin of Gravity

*Fernando Porcelli\* and Giancarlo Scibona*

Department of Environmental Sciences, University of Tuscia, Viterbo, Italy

\*Author for correspondence:

Fernando Porcelli
University of Tuscia
Largo dell'Università
01100 Viterbo ITALY
Phone:+390761357041
e-mail: porcelli@unitus.it

ABSTRACT

The Schwarzschild's black hole dynamics in presence of gravity is described on using the thermodynamic equations of state for contractile materials. Its entropy and temperature, obtained by using classical principles, reproduce the results derived from quantum field theories and statistical mechanics. The given results show that, by using the gravitational dynamics to reproduce the thermodynamic equation $TdS/dx = F_{gravity}$, there is no way to establish the entropic origin of gravity, because the results can be seen the other way around.

INTRODUCTION

In black hole physics, quantum theory, gravitation and thermodynamics are strictly correlated, and this fact indicates that the black hole's dynamics can be identified with the thermodynamic laws. In this paper progress in this direction is reported. In fact, we show (i) that the thermodynamic equation of state for contractile material in closed system accounts for the Schwarzschild's black hole entropy increase due to the gravitational work, and (ii) that its temperature and entropy, obtained from classical principles and thermodynamic laws, agree with the Hawking's temperature and the Bekenstein-Hawking's entropy[1-5], whose microscopic origin, on counting the number of microstates of extremal black holes in string theories[6] and the number of quantum mechanically distinct ways that the hole could be made[7], has been brought back to statistical mechanics. Moreover, (iii) we discuss the euristic hypothesis[8,9] that gravity is an entropic emergent phenomenon and we draw the conclusion that, using the thermodynamic equation $TdS/dx = F_{gravity}$, or its relativistic analogue[9], there is no way to state the gravity entropic origin, because the results can be seen the other way around.

THEORETHICAL BACKGROUND

*Black holes thermodynamics.*

In Black hole physics[1-5] (i) the Kerr-Newman's black hole conservation law $[d(mc^2) = \Theta dA + \Phi dq + \Omega dj]$ is identified with the first thermodynamic law

$$dE = TdS + \Phi dq + \Omega dj, \tag{1}$$

and the work $[\Phi dq + \Omega dj]$ done on the black Hole by an external agent, who increases its charge q and angular momentum *j* by *dq* and *dj*, respectively, is identified with $(-PdV)$ in thermodynamics.

(ii) The black hole's entropy, proportional to the horizon area *A*,

$$S = K_B(c^3/4G\hbar)A \quad (G\hbar/c^3 \text{ Planck length squared}), \tag{2}$$

and the principle that the area cannot decrease[5,10,11] allow to identify the condition $dA \geq 0$ for any change of the black hole, with $dS \geq 0$ for any change of a thermodynamic system.

Moreover, an incipient black hole is in a radiating state[3,4], the emergent radiation has to carry enough entropy to compensate the black hole's mass-entropy loss, and its temperature is the Hawking's temperature[3,4] of the emitted radiation

$$T_H = g\hbar/2\pi c K_B, \quad g = (GM/r^2), \tag{3}$$

a result that is like to the Unruh's[12] temperature ($T_U = a\hbar/2\pi c K_B$) experienced by an observer in a frame accelerated by a given force and that can be expressed in equivalent forms

$$T_H = \hbar c^3/8\pi K_B GM = 2GM\hbar/AcK_B, \quad A = 4\pi r^2, \quad r = 2GM/c^2. \tag{4}$$

Thermodynamics is a physical science of universal content and, then, is not a case that the black holes physics do not violate the thermodynamic laws.

*Temperature and entropy from classical principles.*

In thermodynamics, the energy change for a linear displacement ($\Delta x$) of a particle of mass $m$ accelerated by a force in non relativistic 3-space is established by the equipartion principle

$$K_B T/2 = \Delta E = F\Delta x = ma\Delta x. \tag{5}$$

Hence, using the rationalized acceleration $a/4\pi$, and for $\Delta x$ the Compton length ($\hbar/mc$), the correlation between temperature and acceleration is given by

$$K_B T = a\hbar/2\pi c = a(\hbar/2\pi c), \quad T = a\hbar/2\pi c K_B. \tag{6}$$

Moreover, when a particle accelerated by the gravity becomes trapped on the black hole's surface, its mass-energy increase is equal to the entropy change ($c^2 dM = TdS$). Hence, noting the (4) and [$dA = 4\pi r dr = (32\pi G^2/c^4)MdM$, $r = 2GM/c^2$],

$$dS = c^2 dM/T = (8\pi K_B G/\hbar c)MdM = K_B(c^3/4G\hbar)dA, \tag{7}$$

whose integrals reproduce the Eq. (2)

$$S = K_B 4\pi GM^2/\hbar c = K_B Ac^3/4G\hbar. \qquad (8)$$

In conclusion, the black hole's temperature (Eq.6 with $a = g$) and entropy (Eq.8), obtained from classical principles, reproduce the Hawking's temperature (Eq.3) and the Bekenstein-Hawking entropy (Eq. 2).

*Thermodynamics and gravity*

In thermodynamics, the equations of state for contractile material in closed system

$$dE = TdS - PdV + Fdx. \qquad (9)$$

$$(\partial E/\partial x)_T = T(\partial S/\partial x)_T - p(\partial V/\partial x)_T + F. \qquad (10)$$

have been used to investigate the energy and entropy contributions to the elastic force of rubber and steel materials. Here, we extend their use to the black holes and investigate the mass increase of a Schwarzschild's black hole (specified by only the mass $M$), when an elementary particle of mass $m$, distant one Compton length and considered part of the black hole, shortens its distance from the surface under the gravitational force, and becomes trapped on the surface.

In black holes physics, the horizon is defined by those outgoing light rays that just hover under the strong surface's gravity[13]. For a spherical Schwarzschild's black hole, the value of the coordinate $r$ at the surface is $r = 2GM/c^2$ and the horizon area is defined by a transverse measurement of a particular spherical surface [$A = 4\pi r^2 = 8\pi G^2 M^2/c^4$].

The volume inside a black hole, instead, requires a definition of the particular 3-space in which the volume is computed, and can be time dependent or zero, as in the Schwarzschild's solution, where there is zero volume inside the black hole in any Schwarzschild time slice of a Schwarzschild's black hole spacetime[13], and then $PdV = 0$ in the Eqs.(9,10), as in the case of rubber and steel, where $(\partial V/\partial x)_T \approx 0$ at low stretching. Therefore, noting (i) $(\partial V/\partial x)_T \approx 0$, (ii) the Gibbs equation for closed system [$dG = -SdT + Fdx$], and (iii) its cross relation

$$(\partial F/\partial T)_x = -(\partial S/\partial x)_T, \tag{11}$$

The Eq. (10) becomes

$$(\partial E/\partial x)_T = -T(\partial F/\partial T)_x + F, \tag{12}$$

and then (Eqs. 10,12)

$$(\partial S/\partial x)_T = -F/T, \text{ when } (\partial E/\partial x)_T = 0 \tag{13}$$

$$(\partial E/\partial x)_T = F, \text{ when } (\partial F/\partial T)_x = 0 \tag{14}$$

$$(\partial E/\partial x)_T = 0, \text{ when } F = T(\partial F/\partial T)_x. \tag{15}$$

RESULTS

In the rubber's case, noting that

$$[F = -(\text{constant})K_B Tdx, \; F = T(\partial F/\partial T)_x, \; (\partial E/\partial x)_T = 0, \tag{16}$$

the entropy decreases on stretching [$(\partial S/\partial x)_T < 0$, being $(\partial F/\partial T)_x > 0$, Eq.11] and the Eq. (13) states the entropic origin of the rubber's elastic force. For steel, instead, the entropy does not change [$(\partial S/\partial x)_T = 0$, being $(\partial F/\partial T)_x = 0$, Eq.(11)] and the energy gradient is the major contribution to its elastic force (Eq. 14). These findings agree with the Boltzmann's entropy definition: the rubber's molecules, randomly kinked coils, become oriented on stretching, and then its entropy decreases, while for steel the crystalline order of its structure is hardly affected by the stretching force.

*The black hole case*

In the Schwarzschild's black hole case, when a particle of mass *m* (considered part of the black hole) shortens its distance from the surface by the amount ($-\Delta x = -\hbar/mc$) under the gravitational force ($F_G$), the condition [$(\partial E/\partial x)_T = 0$, Eq.15] holds, because the force $F_G$ depend on the temperature (Eq. 13)

$$F_G = mg = 2\pi K_B T_H (mc/\hbar), \; F_G = T_H(\partial F_G/\partial T_H)_x, \; (\partial E/\partial x)_T = 0, \tag{17}$$

and then the Eq. (13) states the entropic origin of gravity

$$(\partial S/\partial x)_T = F_G/T_H = (GMm/r^2)/T_H = 2\pi K_B(mc/\hbar). \tag{18}$$

However, in black holes, (i) any change due to the gravitational work ($mg\Delta x$) shows itself only in the entropy increase

$$[\Delta S = mg\Delta x/T_H = 2\pi K_B, \quad (\Delta S/\Delta x) = mg/T_H = 2\pi K_B mc/\hbar], \tag{19}$$

(ii) the gravity, already included in the Eqs. (3,18,19) is necessarily reproduced

$$T_H(\Delta S/\Delta x) = T_H[2\pi K_B(mc/\hbar)] = GMm/r^2, \tag{20}$$

and then the given results (Eqs.18-20) do not establish the gravity entropic origin, because can be interpreted the other way around.

In conclusion, the Eqs (13-15) (i) describe the Schwarzschild's black hole dynamics in presence of gravity, (ii) allow to reproduce its entropy (Eqs.7,8) because, once the particle becomes trapped on the horizon, the black hole's mass increase is equal to the entropy change ($T_H \Delta S = c^2 \Delta M$), and (iii) strengthen further the correlation between black hole's physics and thermodynamics, whose laws, on using the non equilibrium thermodynamics principles, describe the rate of black hole's mass-entropy change during its life[14,15].

DISCUSSION AND CONCLUSIONS

The given results (Eqs.18-20) – noting that information on the black hole's internal structure are unavailable - can be interpreted in two opposite ways: gravity increases the entropy and gravity is an entropic force. In fact, in absence of a definite correlation between the entropy increase and the internal structure change of a given system, as in the case of polymer's elastic force, there is no way, on using equations like the (3,19), to state the gravity entropic origin. In fact, the gravitational force, already included in the equations is necessarily reproduced (Eqs.18-20), and then the results can be interpreted the other way around.

*The entropic origin of gravity* [8,9]

In ref.[8], the identity between gravitational field equations and horizon thermodynamics is discussed for a wide class of models, but the interpretations of the mathematical results are only euristic hypothesis. The given results [pp.77-79], in fact, are based on the same postulates: (i) in an emergent spacetime, where quantum and thermal fluctuations are in reciprocal relation, the entropy gradient $\Delta S/\delta x$ arises from the gradient ($\Delta n/\delta x$) in the microscopic degrees of freedom over a region $\delta x$; (ii) in the motion of a particle accelerated by the potential ($V = mgx$), the quantum mechanical averages satisfy the relation ($\delta E = mg\delta x = F\delta x$); (iii) in the local Rindler frame the first law ($\delta E = T\Delta S$) holds.

Hence, on using [$\Delta S = 2\pi K_B$, $\delta x \approx \hbar/mc$], the Rindler temperature [$T = g\hbar/2\pi K_B c$] and the entropy gradient [$\Delta S/\delta x = 2\pi K_B/\delta x = 2\pi K_B mc/\hbar$] like the Eqs. (3,19), gravity is necessarily reproduced [$T\Delta S/\delta x = mg$]. Therefore, noting that gravity is already included in the equations and that in an emergent spacetime, where the microscopic degrees of freedom may be uniformly distributed over a region $\delta x$, their agglomeration due to gravity increases disorder and entropy, the given results do not establish the gravity entropic origin, because can be interpreted the other way around.

The same arguments hold in the case of ref. [9, pp.6-9], where the postulated temperature ($T = 2MG\hbar/AcK_B$) and entropy gradient [$\Delta S/\Delta x = 2\pi K_B(mc/\hbar)$] like the Eqs.(4,19), necessarily reproduce the gravity ($T_H\Delta S/\Delta x = mg$). Moreover [9, p.25], in the context of ADS/CFT (Anti de-Sitter Space/Conformal Field Theory), where a black hole is dual to a thermal state on the boundary and the particle is represented as a delocalized operator that is gradually thermalized, it is argued that, when a particle reaches the horizon and becomes part of the thermal state, gravity becomes an entropic force near the horizon, because the black hole's mass-entropy changes is equal to the gravitational work [$T_H\Delta S = c^2\Delta M = mg\Delta x$], and then ($T_H\Delta S/\Delta x = mg$). The same result as above that does not establish the gravity entropic origin, because can be seen the other way around.

In the context of microscopic theories [9, pp.17,18], it is stated that gravity is an entropic force. However, this statement is only one of the possible interpretations of the given results. In fact, the temperature [$T = \hbar/2\pi(e^\phi N^b \nabla_b \phi)$, with $N^b$ a unit outward pointing vector normal to the holographic screen surface and to the Killing vector $\xi^b$, $e^\phi$ redshift factor, $\phi$ Newton's potential] and the entropy gradient [$\nabla_a S = -2\pi(m/\hbar)N_a$] are the relativistic analogues of the Eqs (3,19) and, then, the gravity relativistic analogue [$T\nabla_a S = -me^\phi \nabla_a \phi$] is necessarily reproduced. These results, once again, allow a twofold interpretation: gravity is an entropic force; gravity increases disorder and entropy.

In conclusion, gravity is a fundamental force and, in absence of a definite correlation between entropy and internal structure changes of a given system, as in the case of polymer's elastic force, any hypothesis on the gravity entropic origin, based on equations like the (3,19), is only one of the possible interpretations, because one can also interpret the results the other way around: the entropy increase arises from the gravitational work.